\documentclass[runningheads]{llncs}

\usepackage{cite}
\usepackage{amsmath,amssymb,amsfonts}
\usepackage{algorithmic}
\usepackage{graphicx}
\usepackage{textcomp}
\usepackage{bmpsize}
\usepackage[dvipsnames]{xcolor}
\usepackage{lipsum}
\usepackage{glossaries}
\usepackage{tikz}
\usepackage{tabularx}
\usepackage{booktabs}
\usepackage{microtype}
\usepackage{listings}
\usepackage{hyperref}
\usepackage{cleveref}
\usepackage{balance}
\usepackage{csquotes}
\usepackage{adjustbox}
\usepackage{array}
\usepackage{textcomp}
\usepackage[breakwithin]{parnotes}
\usepackage{todonotes}
\usepackage{diagbox}
\def\BibTeX{{\rm B\kern-.05em{\sc i\kern-.025em b}\kern-.08em
T\kern-.1667em\lower.7ex\hbox{E}\kern-.125emX}}

\newacronym[firstplural=Abstract Syntax Trees (AST)]{ast}{AST}{Abstract Syntax Tree}
\newacronym{pdg}{PDG}{Program Dependence Graph}
\newacronym{mcl}{MCL}{Markov Cluster Algorithm}
\newacronym{pypi}{PyPI}{Python Package Index}
\newacronym{dbscan}{DBSCAN}{Density-Based Spatial Clustering of Applications with Noise}
\newacronym{hdbscan}{HDBSCAN}{Hierarchical Density-Based Spatial Clustering of Applications with Noise}
\newacronym{acme}{ACME}{AST Clustering using MCL to mimic Expertise}

\newcommand{\mcx}[1]{\multicolumn{3}{>{\hsize=\dimexpr2\hsize+3\tabcolsep\relax}c}{#1}}

\newcolumntype{R}[2]{%
  >{\adjustbox{angle=#1,lap=\width-(#2)}\bgroup}%
  l%
  <{\egroup}%
}

\lstdefinelanguage{JavaScript}{
  morekeywords=[1]{break, continue, delete, else, for, function, if, in,
      new, return, this, typeof, var, void, while, with},
  morekeywords=[2]{false, null, true, boolean, number, undefined,
      Array, Boolean, Date, Math, Number, String, Object},
  morekeywords=[3]{eval, parseInt, parseFloat, escape, unescape},
  sensitive,
  morecomment=[s]{/*}{*/},
  morecomment=[l]//,
  morecomment=[s]{/**}{*/}, 
  morestring=[b]',
  morestring=[b]"
}[keywords, comments, strings]

\begin{document}

\title{Supporting the Detection of Software Supply Chain Attacks through Unsupervised Signature Generation}
\titlerunning{Supporting the Detection of Software Supply Chain Attacks}


\author{Marc Ohm\inst{1} \and
  Lukas Kempf\inst{1,2} \and
  Felix Boes\inst{1} \and
  Michael Meier\inst{1,3}
}

\institute{University of Bonn, Institute for Computer Science 4,\\ Friedrich-Hirzebruch-Allee 8, 53115 Bonn, Germany\\
  \email{\{ohm,boes,mm\}@cs.uni-bonn.de} \and
  Hochschule Bonn-Rhein-Sieg, Institut für Sicherheitsforschung,\\ Rathausallee 10, 53757 Sankt Augustin, Germany\\
  \email{kempf@uni-bonn.de}\\\and
  Fraunhofer FKIE, Department for Cyber Security,\\ Zanderstraße 5, 53177 Bonn, Germany}

\authorrunning{M. Ohm et al.}

\maketitle

\begin{abstract}
  Trojanized software packages used in software supply chain attacks constitute an emerging threat.
  Unfortunately, there is still a lack of scalable approaches that allow automated and timely detection of malicious software packages and thus most detections are based on manual labor and expertise.
  However, it has been observed that most attack campaigns comprise multiple packages that share the same or similar malicious code.
  We leverage that fact to automatically reproduce manually identified clusters of known malicious packages that have been used in real world attacks, thus, reducing the need for expert knowledge and manual inspection.
  Our approach, \glsfirst{acme}, yields promising results with a $F_{1}$ score of 0.99.
  Signatures are automatically generated based on characteristic code fragments from clusters and are subsequently used to scan the whole npm registry for unreported malicious packages.
  We are able to identify and report six malicious packages that have been removed from npm consequentially.
  Therefore, our approach can support analysts by reducing manual labor and hence may be employed to timely detect possible software supply chain attacks.


\end{abstract}

\glsresetall

\section{Introduction}
  Modern software is developed on top of an ever-growing pool of third party tools, libraries and packages.
Hence, the integrity of software projects depend directly on the integrity of the underlying software supply chain.
Over the past few years, software supply chain attacks that leverage trojanized software packages kept emerging~\cite{ohm2020backstabber}.
A central role in that ecosystem is held by maintainers of package repositories like npm or \gls{pypi}.
These platforms are repeatedly abused for the distribution of trojanized software packages that are part of a software supply chain attack.

From the point of view of a malware author, distributing trojanized software packages to open source package repositories is a both cost-efficient and effective for a number of reasons.
Here, we name a few.
\begin{itemize}
  \item Open source packages are frequently used in various software projects.
        However, the packages code base is unlikely to be audited by the user of the package.
  \item Most package repositories do not employ effective malware checks.
        Therefore, malicious packages tend to be available for roughly 200 days before being identified and removed~\cite{ohm2020backstabber}.
  \item Moreover, the lack of effective malware checks allows malware authors to republish slightly modified malicious code fragments without beeing detected timely~\cite{ohm2020backstabber}.
\end{itemize}

In a package repository, malicious packages are most likely to be found by security analysts of the repository owner, by a third party like
Sonatype or by individuals.
Schematically, analysts operate as follows.
\begin{enumerate}
  \item An analyst is informed of the presence of a suspicious packages.
  \item The analyst works through all code files to find characteristic code fragments of a trojanized package.
  \item The characteristic code fragments of the recent trojanized package are compared with characteristic code fragments of malicious packages found in the past.
        Forming malware families, this leads to a clustering of the trojanized packages.
\end{enumerate}
This procedure comprises three problems.
To begin with, an analyst needs to get informed about new suspicious package.
Then, the analyst needs to recall all previously used code fragments to find similarities.
Lastly, since malware authors regularly republish variations of their malicious code fragments,
it is mandatory to periodically search for said characteristic code fragments in the whole repository to keep the repository clean.

%

Since identifying malicious packages is time-consuming and comprises tedious steps, malicious packages tend to be available from package repositories for roughly 200 days.~\cite{ohm2020backstabber}
Clearly, an improvement of automated capabilities for timely detection of such attacks is mandatory.
The earliest point in the lifecycle of a software package, at which a third party gets access to the source code, is when a maintainer uploads it to a package repository.
As said, package repositories play a central and critical role in the software supply chain ecosystem.
Thus, maintainers of package repositories should ensure the quality and integrity of uploaded packages.
This has been acknowledged and implemented to some extent.
For instance \gls{pypi} performs Malware Checks\footnote{\url{https://warehouse.pypa.io/development/malware-checks.html}}.
However, they are implemented rather rudimentary.

In this paper, we provide a tool that automatizes the most time-consuming and tedious tasks of the security analyst.
Hereby, our tool is solely code driven and can be operated by repositories' owners as well as by a third party.

Based on a manual annotated dataset, we evaluate various automated approaches to mimic the manual clustering of malicious packages by an expert.
Following the idiom \textquote{if you've seen one, you've seen them all} we automatize the identification of related malicious packages to keep the upper hand in the arms race of software supply chain attacks.
Consequentially, we propose a timely detection of malicious packages based on signatures derived from identified clusters.
To this end, we leverage \glspl{ast} that are generated from known malicious packages that have previously been used in real world attacks.
Eventually, clusters of packages that share source code are identified through \gls{mcl}.
Our results indicate excellent performance ($F_{1}=0.99$) on the annotated dataset and good scalability for practical application.
This way, suspicious packages are detected as soon as they are published to a package repository.
It supports an analyst by notifying him about suspicious packages and giving hints about possible connections to already known malicious packages.

The contribution of this paper is the automated and signature-based detection of possibly malicious packages which may then be analyzed by an expert.
Furthermore, these signatures are generated automatically and unsupervised which further reduces manual work.
Thus, the approach is suited for an early detection of trojanized software packages.
It may be employed by maintainers of package repositories in order to stop software supply chain attacks by removing malicious packages before they are distributed.
Eventually, we were able to detect and report six incidents of malicious packages on npm that share code with previously distributed malicious packages.

The remainder of this paper is organized as follows.
\Cref{sec:related-work} provides related work to frame the academic context of our approach.
The underlying methodology for our approach is depicted in \Cref{sec:methodology}.
In \Cref{sec:detail} necessary backgrounds for this work are presented.
Our results are presented in \Cref{sec:results} and subsequently discussed in \Cref{sec:discussion}.
\Cref{sec:conclusion} concludes the paper and provides an outlook for future work.

\section{Related Work}\label{sec:related-work}
  The detection of similar code fragments is being used for the detection of software plagiarism.
To this end, a vast majority of approaches leverage \glspl{ast}~\cite{chilowicz2009syntax,djuric2013source,novak2019source,cosma2011approach,ragkhitwetsagul2018comparison}.

The detection of source code similarity also found application in cyber security.
Especially, the detection of software vulnerabilities leverages code similarity detection in order to identify known but currently undetected vulnerabilities in software.
Known vulnerable source code is used to generate signatures which are used to scan other source codes for these known patterns~\cite{chinthanet2020code,bilgin2020vulnerability,li2016vulpecker,yamaguchi2012generalized}.

In contrast to vulnerable software components, the research field of malicious software packages is comparably new.
Most often they try to identify possible typosquatting attacks~\cite{tschacher2016typosquatting,vu2020typosquatting,taylor2020spellbound}.
Nonetheless, there are approaches that try to detect malicious packages based on their source code.
For instance by looking for anomalies in a package's source codes compared to all other packages~\cite{carnogursky2019attacks,garrett2019detecting} or by leveraging heuristics~\cite{pfretzschner2017identification,duan2020measuring} of presumably malicious characteristics.
Moreover, dynamic analysis of suspicious packages might be considered in order to detect malicious behavior~\cite{ohm2020towards, duan2020measuring}.
Furthermore, Ohm et al.\@~\cite{ohm2020backstabber} systematized software supply chain attacks by collecting and analyzing a large dataset of malicious open source packages that have been used in real world attacks.

Now that an annotated dataset of known malicious packages is at hand, it is possible to leverage insights and ground truth to develop suitable detection techniques.
Our contribution differs from the approaches mentioned beforehand by leveraging evidence-based characteristics of known malicious packages.
In this work, we focus on static code analysis in order to detect malicious packages based on automatically generated signatures.
Besides \glspl{ast} that are widely used in related work we evaluate text-based approaches and \glspl{pdg}.

\section{Methodology}\label{sec:methodology}
  As observed by Ohm et al.\@~\cite{ohm2020backstabber}, malicious packages tend to have characteristic code fragments in common.
This might be because they are employed in the same attack campaign or were simply copied by other malware authors.
As discussed in the introduction, an automated approach that solves the time-consuming and tedious tasks to identify and search for said characteristic code fragments is highly anticipated.
In this paper, we present such a tool by mimicking the analysts approach.
More precisely, we provide and evaluate several automated approaches to
(1) automate the clustering of (potentially malicious) software packages using characteristic code fragments and
(2) leverage results for the detection of currently unidentified clones of known malicious packages.

To begin with, we model the analysts approach (that is partially mimicked by our tool) as follows.
\begin{enumerate}
  \item An analyst is informed of the presence of a suspicious package.
  \item The analyst works through all code files to find characteristic code fragments of a trojanized package.
  \item \label{task:cluster} The characteristic code fragments of the recent trojanized package are compared with characteristic code fragments of malicious packages found in the past.
        Forming malware families, this leads to a clustering of the trojanized packages.
  \item \label{task:signatures} Having sophisticated knowledge of the malware clusters, the analyst scans the full package repository to detect new suspicious packages.
\end{enumerate}

\begin{figure}[tb]
  \includegraphics[width=\columnwidth]{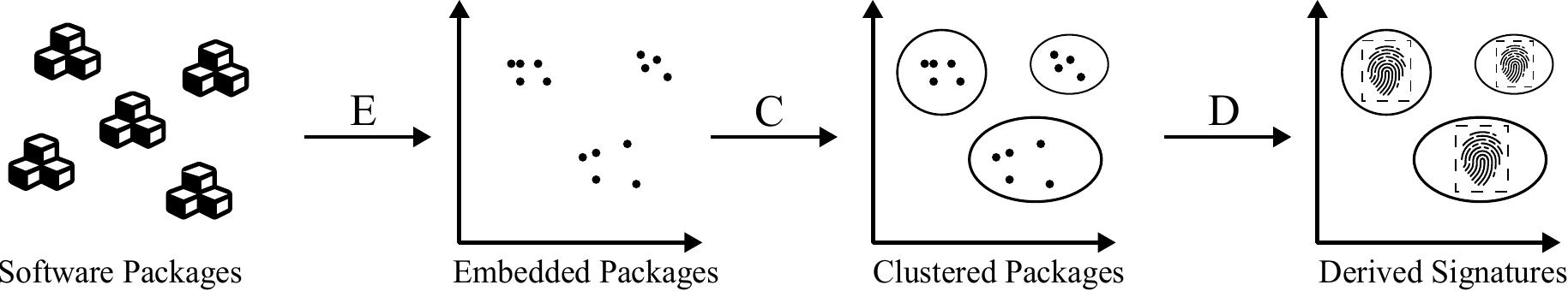}
  \caption{Steps performed to generate signatures for software packages. First, software packages are embedded into a metric space. Clusters are identified based on their distance. Finally, a signature is derived for each cluster.}\label{fig:ablauf}
\end{figure}

To automate the most tedious and time-consuming tasks \ref{task:cluster} and \ref{task:signatures}, we proceed as depicted in \Cref{fig:ablauf}.
In the first step, we embed each package into a metric space, i.e., each package is represented by a point (or even a cloud of points) and each pair of points has a precise distance.
Then, using the distances of the points to each other, clusters are computed.
Hereby, various methods to embed and cluster the software packages are evaluated against the manual clustering of an expert (on an annotated data set).
Then, each cluster represents a family of malicious packages that share characteristic malicious code fragments.
From each cluster, the set of characteristic code fragments is extracted and used to derive a signature of the cluster.

After providing our metric that measures how well the automated clustering mimics the manual approach, we evaluate the clustering on an annotated dataset.
In order to show that our tool is efficient and feasible for practical application on the large scale, it is evaluated on the full npm repository.
Hereby, we are able to identify and report six malicious packages that have been removed from npm consequentially.

\subsection{Embedding of Packages}\label{sec:meth:embedding}
  Each package consists of a number of source files and each source file consists of a number of functions.
  In our approach, a package is associated either with its source files or with its functions.
  More precisely, each function is represented either as Program Text, Program Dependence Graph or Abstract Syntax Tree.
  Hereby, we treat member functions as independent functions and group all statements in the global scope into a function.
  Then, the distance between two packages is computed via string similarity or graph similarity.
  The details are found in \Cref{sec:details:embedding}.

\subsection{Clustering of Packages}\label{sec:meth:mimic_clustering}
  The defined goal is an automated approach that mimics the experts clustering.
  To this end, we evaluate several unsupervised clustering algorithms on the packages (using the distances between their representations) and compare the resulting clusters with the experts clustering of malicious packages.
  Hereby, unsupervised methods have been chosen to reduce the need for prior knowledge about the dataset.
  In this section, we provide our metric to compare two clusterings and discuss the annotated dataset used in the evaluation.
  In \Cref{sec:details:clusterin}, (the details of) the evaluated clustering algorithms are presented.

  All approaches are evaluated on the \textquote{Backstabber's Knife Collection} dataset~\cite{ohm2020backstabber}\footnote{For reproducibility we use the version of the dataset as described in the publication.} as ground truth.
  This dataset contains malicious packages harvested from npm, \gls{pypi}, and RubyGems as well as a clustering of these which was performed manually and with expert knowledge.
  For sake of brevity and with respect to the amount of npm packages in the dataset, we focus on packages written for Node.js in JavaScript.
  However, our approach is transferable to all kind of programming languages.

  In order to compare the manual clustering by Ohm et al.\@~\cite{ohm2020backstabber} with a fixed automated clustering approach, the conceiving metrics Precision, Recall and $F_1$-score are employed as follows.
  First of all, we assume that the manual clustering is complete and accurate, i.e., every malicious code similarity is found and packages are clustered correctly.
  Then, a pair of packages is said to be a
  \begin{itemize}
    \item \textbf{true positive} if the two packages are in the same cluster in both approaches.
    \item \textbf{true negative} if the two packages are in different clusters in both approaches.
    \item \textbf{false positive} if the two packages are in different manually generated clusters but in the same automatically generated cluster.
    \item \textbf{false negative} if the two packages are in the same manually generated clusters but in the different automatically generated clusters.
  \end{itemize}
  With this definition, the metrics Precision, Recall and $F_1$-score are interpreted as follows.
  By definition, Precision is the ratio of true positives in all positives.
  Therefore, the Precision is high if the number of false negatives is relatively low.
  Observe that this is the case if the automated approach generates clusters that are overall finer or as fine as the manual approach.
  Observe analogously that the Recall is high if the automated approach generates clusters that are overall coarser or as coarse as the manual approach.
  Consequentially, the $F_1$-score (which is the harmonic mean of Precision and Recall) measures how well the automated clustering mimics the manual approach.

\subsection{Derivation of Signatures}\label{sec:meth:signatures}
  It turns out that \gls{acme} mimics the manual clustering of an expert nearly perfectly with an $F_1$-score of $0.99$.
  Based on top of \gls{acme}, the signature of a cluster of malicious packages is derived by constructing a so called fingerprint from each characteristic code fragment.
  Here, we explain what a characteristic code fragment is.
  Fingerprinting is based on Chilowicz et al.\@~\cite{chilowicz2009syntax} and is explained in detail in \Cref{sec:details:fingerprinting}.
  In \Cref{sec:signatures} and \Cref{sec:npm-scan}, we demonstrate that these automatically generated signatures are quite good and yield high quality signatures with just a few minutes of manual refinement.

  For each cluster $c$ one or more characteristic code fragment are chosen to serve as signature $S_c$ for that cluster as follows.
  By construction, a cluster consists of a set of ASTs and to each AST, we associate a fingerprint $\mathcal H$ based on an approach by Chilowicz et al.\@~\cite{chilowicz2009syntax}.
  Roughly speaking, a fingerprint is simply a hash of a given AST, see \Cref{sec:details:fingerprinting} for details.
  Now a fingerprint $\mathcal H$ is ``characteristic'' if the following conditions are met.
  \begin{enumerate}
    \item \label{relevant_fp:unique} The code fragment $\mathcal H$ is unique to its cluster, i.e., $\mathcal H$ is not derived from any package in any other cluster.
    \item \label{relevant_fp:recurring} The code fragment $\mathcal H$ is derived from at least two packages in its cluster.
    \item \label{relevant_fp:not_basic} The code fragment $\mathcal H$ cannot be derived from one of the 108 most depended upon packages\footnote{\url{https://www.npmjs.com/browse/depended}, we are limited to the 108 most depended upon packages due to technical issues of the website.} from npm.
  \end{enumerate}
  Observe that \ref{relevant_fp:unique}. ensures that the signatures of the clusters are pairwise disjoint.
  This means that a newly classified package is assigned to one cluster only.
  Observe further that the condition \ref{relevant_fp:recurring}. focuses signatures on common and hence descriptive code for the analyzed cluster.
  Condition \ref{relevant_fp:not_basic}. forbids code fragments that can be found in popular and hence supposedly benign packages.

\subsection{Measuring Effectivity and Efficiency on the Large Scale}\label{sec:meth:large_scale}
  In order to show that our approach is efficient and feasible for practical application on the large scale, it is evaluated on the full npm repository in \Cref{sec:npm-scan}.
  As a result, we identified and report six malicious packages that have been removed from npm consequentially.
  Furthermore, a theoretical inspection of the efficiency by means of runtime and space complexity is carried out in \Cref{sec:disc:efficiency}.

\section{Details}\label{sec:detail}
  In this work, we identify syntactic clones as well as variations of malicious packages.
This requires a technique that is able to represent code fragments with a certain amount of abstractness and a clustering approach to pair similar code fragments on the chosen abstract representation.
In this section, we present and discuss the details of our approach introduced in \Cref{sec:methodology}.

\subsection{Syntactic Clones}
  Walker et al.\@~\cite{walker2020open} define four clone types which are related in varying degrees.
  Clones of Type-1 share two identical code fragments without respect to whitespace, blanks, or comments.
  If the structure of the code is the same but some functions, classes, or variables are renamed, we speak of Type-2 clones.
  Type-3 clones differ in naming but also show differences in structure, i.e., some code fragments may be modified.
  If no syntactical similarity can be observed but the function is the same, we speak of Type-4 clones.

\subsection{Embedding of Package}\label{sec:details:embedding}
  In order to identify code clones we need to compare code fragments.
  For that purpose we evaluate several approaches, namely we compute the similarity via string similarity~\cite{ragkhitwetsagul2018comparison}, \glsfirst{pdg}~\cite{liu2006gplag}, and \glsfirst{ast}~\cite{vandongen2000cluster}.

  For text-based similarity the Python package \texttt{FuzzyWuzzy}~\cite{seargeek11fuzzywuzzy} is employed.
  It offers multiple modi but with respect to the work of Ragkhitwetsagul et al.\@~\cite{ragkhitwetsagul2018comparison} solely simple\_ratio, partial\_ratio, token\_sort\_ratio, and token\_set\_ratio are evaluated.
  All of these modi leverage the Levenshtein distance~\cite{kruskal1983overview} to calculate the difference between two inputs.
  In our case, we extract the code fragments of all functions of the package.
  The similarity of two packages is the smallest Levenshtein distance found between two code fragments of the two packages.
  However, text-based approaches performed worse than \gls{ast} thoroughly and are thus not employed.

  We implemented \gls{pdg} according to the description of Liu et al.\@~\cite{liu2006gplag}.
  However, in the generation of \gls{pdg} for the malicious packages, we observed disproportionate runtime in combination with performance below average.
  Thus, we discarded \gls{pdg} for further experiments.

  AcornJs~\cite{haverbeke2020acorn}, a lightweight parser for JavaScript, is used to transform source code into \gls{ast} representation.
  Comparison of multiple \glspl{ast} allows focusing on the identification of structural similarities.
  Through abstraction of source code into a structural representation naming of identifiers is of no matter.

\begin{figure}[tb]
    \begin{lstlisting}
  function fib(n) {
      if (n < 1) return 0;
      else if (n <= 2) return 1;
      return fib(n-1) + fin(n-2);
  }
      \end{lstlisting}
    \includegraphics[width=\columnwidth]{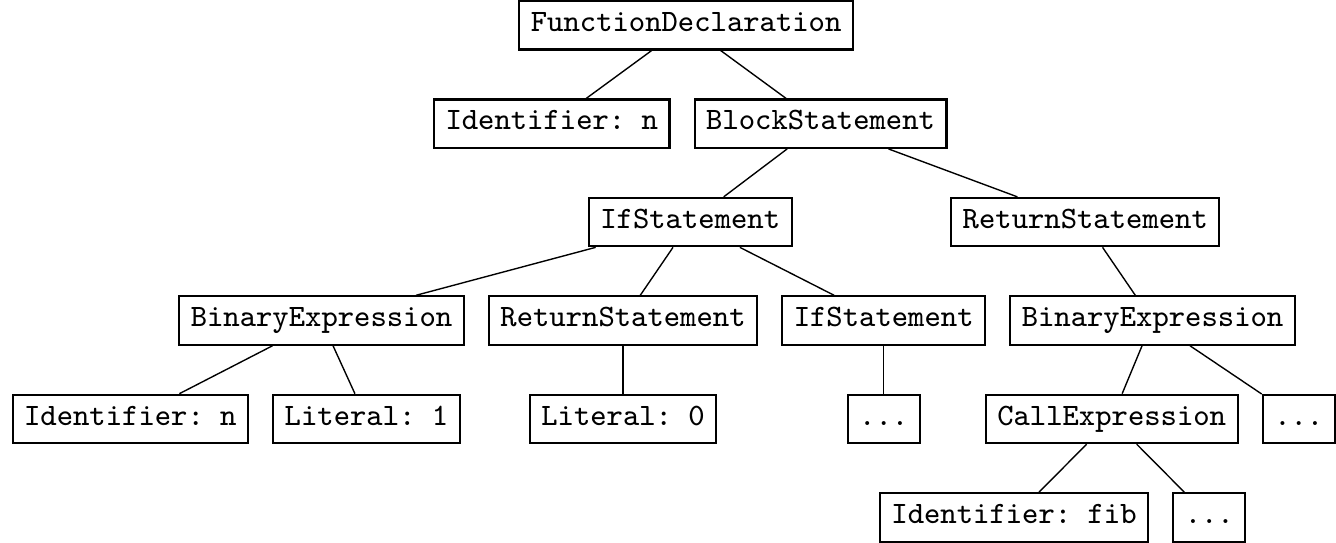}
    \caption{Example of an \gls{ast}.}\label{fig:ast}
  \end{figure}

  As visualized in \Cref{fig:ast}, the \gls{ast} of a code fragment is a tree that represents every structural element of the code as node.
  The tree starts with the function declaration as root node.
  On the left side of the root the function parameters are enumerated (only \texttt{n} in this case).
  To the right, the body of the function is unfolded.
  It comprises the if-else-statement as well as the return statement.
  This way a tree representation of the source code is generated.
  Because of syntactic differences in programming languages, \glspl{ast} are language-dependent.

  Similarity of two \glspl{ast}, i.e., two packages, is calculated according to the Tree Edit Distance introduced by Zhang-Shasha~\cite{zhang1989simple} using the Python package \texttt{zss}~\cite{henderson2013zhang}.
  The similarity of two packages, A and B, is the smallest Tree Edit Distance when comparing all \glspl{ast}, i.e., all functions, of package A to all \glspl{ast} of package B.
  This allows the detection of Type-1, Type-2, and Type-3 clones.
  Furthermore, way we can quantify the similarity between all malicious packages at hand which is then used to identify clusters among these.

\subsection{Clustering of Packages}\label{sec:details:clusterin}
  After computing the similarities of all pairs of packages, we evaluate the quality of several clustering approaches.
  We evaluate connected component (\textit{ccomp}) and maximal cliques (\textit{clique}) by leveraging the Python package \texttt{NetworkX}~\cite{hagberg2008exploring}.
  \gls{dbscan} is implemented by using the Python package \texttt{scikit-Learn}~\cite{pedregosa2011scikit} and \gls{hdbscan} by \texttt{hdbscan}~\cite{mcinnes2017hdbscan}.
  Last, we examine \glsfirst{mcl}~\cite{vandongen2000cluster} for which we leverage the Python package \texttt{Markov-Clustering}~\cite{allard20markov}.

  \gls{mcl} is an efficient clustering algorithm based on Markov chains for graphs~\cite{vandongen2000cluster}.
  In contrast to approaches like K-Means no a priori knowledge about the underlying data, e.g., number of clusters, is required.

  Roughly speaking, \gls{mcl} performs a random walk of length $k$ through the graph along the edges starting at node $v$.
  For a sparse graph, the random walk reaches some node $w$ in a denser region with a high probability.
  This leverages the characteristics of highly connected sub graphs which have many inner nodes but only few outer nodes.
  However, \gls{mcl} does not walk randomly on the nodes but calculates the probability to reach node $w$ from node $v$ slightly differently.

  The calculation is performed over multiple iterations simultaneous for all nodes and comprises two steps.
  At first, an \textit{expansion} is performed in which all reachable nodes from a starting node $v$ are added to $v$'s matrix of probable neighbors.
  In the \textit{inflation} step, neighboring nodes with a high probability to reach are boosted and weak links are discarded.
  The algorithm terminates when convergence, i.e., no changes to the last iteration, is detected.
  For more details, we refer the reader to \cite{vandongen2000cluster}.

\subsection{Derivation of Signatures}\label{sec:details:fingerprinting}
  Our approach leverages fingerprinting as proposed by Chilowicz et al.\@~\cite{chilowicz2009syntax}.
  From each function $f$ represented in an \gls{ast} $G$, we derive a so called fingerprint $\mathcal H_f$.
  To this end, we focus our attention to the subgraph $G_f \subset G$ associated to $f$ after all (nested) functions that are defined inside $f$ are discarded.
  For each node $v \in G_f$, we concentrate on its type $t(v)$.
  For the example, in \Cref{fig:ast}, each \texttt{Identifier} and the value of each \texttt{Literal} is discarded.
  After fixing SHA-256 as hash function $C$, the fingerprint of $f$ is defined recursively as follows.
  Given an arbitrary node $v \in G_f$ with children $w_1, w_2, \dots$, we define $\mathcal H(v)$:
  \begin{equation}
    \label{eqn:fingerprint_snipped}
    \mathcal H(v) = C\left( t(v) ~\|~ \mathcal{H}(w_1) ~\|~ \mathcal{H}(w_2) ~\|~ \dots  \right)
  \end{equation}
  Denoting the root of $G_f$ by $r_f$, the fingerprint of $f$ is
  \begin{equation}
    \label{eqn:fingerprint}
    \mathcal H_f = \mathcal H(r_f)
  \end{equation}

  We remark that we leverage a different function $t$ than Chilowicz et al.
  Our $t(v)$ solely takes the type of node into account.
  Thus, our subgraph $G_f$ is very focused on the structure of the code fragment by discarding nonstructural information like operators.
  For instance the code fragments $a + b$ and $a * b$ result in the same fingerprint.
  However, $a + (a + a)$ and $(a + a) + a$ yield different fingerprints.
  Let us remark further that we grouped code into a dummy global function if it resides outside of functions, i.e., in global scope.
  Furthermore, we treated functions inside classes as independent functions.

  The signature of a cluster $c$ comprises all characteristic code fragments, i.e., fingerprints, of that cluster (c.f. \Cref{sec:methodology}).
  Now, a package $p$ matches the signature $\mathcal{S}_c$ of cluster $c$ if at least one of $p$'s fingerprints $h^p_1, \ldots, h^p_N$ matches a fingerprint $h \in \mathcal{S}_c$.
  \begin{align}
    \label{eqt:signature_match}
    \mathcal{M}atch_{p,S_c} = \begin{cases}True & \text{if  $h^p_i \in \mathcal{S}_c$ for some $i$}\\ False & \text{else}\end{cases}
  \end{align}

\section{Results}\label{sec:results}
  This section summarizes our results from experiments as introduced in the previous sections.
First, we evaluate which clustering approach is suited best in combination with \gls{ast} to reproduce the results of the manual clustering in \Cref{sec:reproduction}.
The best approach is leveraged in \Cref{sec:signatures} to automatically generate and optimize signatures based on identified clusters.
These signatures are subsequently used in \Cref{sec:npm-scan} to scan the whole npm registry for unreported malicious packages that have code fragments common to known malicious packages.

\subsection{Reproduction of Clustering}\label{sec:reproduction}
  Recall from \Cref{sec:methodology} that we aim to automate the tedious and time-consuming task of manually finding (variations of) recognized malicious code blocks in a given package repository.
  Hereby, packages with similar malicious code blocks are clustered using various unsupervised cluster algorithms.
  In this subsection, we evaluate the quality of these approaches that attempt to reproduce the result of the manual clustering of Ohm et al.\@~\cite{ohm2020backstabber}.

  At the time of evaluation, the dataset used as ground truth contained 114 packages from npm from which 104 packages belong to a cluster.
  Ohm et al. suspected that these packets originated from the same attack campaign~\cite{ohm2020backstabber}.
  If an average attack campaign leverages multiple trojanized packages an automated detection of the undiscovered but related packages is of desire.
  Thus, we propose to automatically generate signatures of known malicious packages in order to support analysts in their task to identify similar packages of the same attack campaign.

  \begin{table*}[tb]
    \caption{Results of \gls{ast} and all clustering approaches with employed parameters, sorted by $F_{1}$ score.}\label{tab:results-all}
    \begin{tabularx}{\textwidth}{lXlXrXrXr}
      \toprule
      Clustering    &  & Parameter                        &  & Precision &  & Recall &  & $F_{1}$ \\
      \midrule
      \gls{mcl}     &  & $\text{exp}=2, \text{inf}=2$     &  & 0.9747    &  & 0.9958 &  & 0.9851  \\
      ccomp         &  &                                  &  & 0.6761    &  & 0.9958 &  & 0.8054  \\
      \gls{dbscan}  &  & $\varepsilon=1, \text{minPts}=2$ &  & 0.6761    &  & 0.9958 &  & 0.8054  \\
      \gls{hdbscan} &  & $\text{minClst}=2$               &  & 0.6580    &  & 0.9967 &  & 0.7927  \\
      clique        &  &                                  &  & 0.9878    &  & 0.6074 &  & 0.7522  \\
      \bottomrule
    \end{tabularx}
  \end{table*}

  In \Cref{sec:meth:mimic_clustering}, the interpretation of Precision, Recall and $F_1$-score is given:
  We achieve a high Precision if the automated approach generates clusters that are overall finer or as fine as the manual approach.
  The Recall is high if the automated approach generates clusters that are overall coarser or as coarse as the manual approach.
  Therefore, the $F_1$-score measures how well the automated clustering mimics the manual approach.
  Thus, it quantifies an approach's ability to find the same clusters unsupervised as an expert would by hand.


  \Cref{tab:results-all} displays the performance of \glspl{ast} in conjunction with all clustering algorithms.
  It is noticeable that most clustering algorithms yield either high Precision or high Recall.
  Solely, \gls{mcl} is capable of reaching both high Precision and high Recall thus recreating the manual cluster as similar as possible.
  With a Precision of 0.97 and a Recall of 1.00 the $F_{1}$ score is at 0.99.
  Through the use of \gls{acme} we are able to recreate the manual clustering performed by expert almost perfectly.

  Having a suitable, automated, and unsupervised clustering at hand, signatures can now be derived to describe the syntactic characteristics of malicious packages of that cluster.
  Using these signatures, packages related to the same attack campaign, i.e., sharing some code fragments, are detected and identified automatically.
  The next chapter will evaluate the quality of the automatically generated signatures to quantify their practical suitability.

\subsection{Quality of Signatures}\label{sec:signatures}
  The previous subsection showed that the experts task of manually clustering malicious code is automated almost perfectly by combining \glspl{ast} with \gls{mcl}.
  Using the \gls{acme} approach described in \Cref{sec:meth:mimic_clustering}, a signature $S_c$ is derived for each cluster $c$.
  Recall from \Cref{sec:meth:signatures} that a signature is a set of relevant fingerprints.
  Recall further that a fingerprint is relevant if
  it is unique to its cluster (condition \ref{relevant_fp:unique}),
  occurs at least twice in its cluster (condition \ref{relevant_fp:recurring}) and
  is not derived from a fixed family of very popular benign packages (condition \ref{relevant_fp:not_basic}).

  In this subsection, we discuss the sizes of the clusters, and we demonstrate that the first two conditions yield signatures with a promising Recall.
  However, the signatures are to coarse, i.e., they produce a huge number of false positives.
  The third condition is mandatory to reduce the number of false positives.
  The quality of the signatures is further improved in the next section.

  In \Cref{tab:cluster-sizes}, the resulting clusters, their sizes, and corresponding number of signatures are shown.
  Our approach automatically identified seven clusters that cover 97 packages.
  As stated initially, 104 packages belong to a manual created cluster in the dataset.
  However, the manual clustering by Ohm et al.\ also took dependency into account for clustering.
  Our approach solely relies on code syntax similarity and hence may not cluster all packages as in the dataset.

  It is noticeable that the sizes of clusters varies heavily ($\sigma^2=230.4082$).
  The smallest ones comprise two packages while the biggest clusters are of size 38 and 36 respectively.
  The size of a signature weakly correlates with the size of the corresponding cluster (Pearson $r=0.65$, $p=0.12$).
  However, there are outliers.
  For instance cluster 1 and 5 yield very large signature compared to their size and in contrast to that cluster 2 yields a very small signature.

  In order to demonstrate that condition \ref{relevant_fp:not_basic}) is mandatory, we test the quality of the signatures associated to all fingerprints satisfying only conditions \ref{relevant_fp:unique} and \ref{relevant_fp:recurring} as follows.
  In a 10-fold cross validation, we cluster the 114 packages containing malicious code with \gls{acme} and derive the signatures associated to all fingerprints satisfying conditions \ref{relevant_fp:unique} and \ref{relevant_fp:recurring}.

  These signatures are evaluated against the 10\% of the split in the cross validation and against 108 benign packages.
  In this context, a package is positive if the automatically generated signature matches the package.
  On average, the Recall is 0.88 but the number of false positives is 46\%.
  This is because the signatures contain too many fingerprints of benign functions, i.e., condition \ref{relevant_fp:not_basic}. is mandatory to reduce the number of false positives.

  In total 3,875 fingerprints satisfying only conditions \ref{relevant_fp:unique}. and \ref{relevant_fp:recurring}. are derived from the malicious packages of the seven clusters.
  Considering relevant fingerprints, i.e., after applying condition \ref{relevant_fp:not_basic}, the seven clusters yield 3,396 (-12.36\%) fingerprints in total.
  These signatures are a good automated approximation that need only a few minutes manual refinement, see \Cref{sec:npm-scan}.
  \begin{table}[tb]
    \caption{Identified clusters based on \gls{acme} sorted by size.
      The size of corresponding signatures $\mathcal S_c$ and matches $M_c$ are based on the level of optimization: Consider all fingerprints satisfying conditions \ref{relevant_fp:unique} and \ref{relevant_fp:recurring}, all characteristic fingerprints, and all characteristic fingerprints with manual optimization.}\label{tab:cluster-sizes}
    \begin{tabularx}{\columnwidth}{Xr|rXr|rXr|rXr}
      \toprule
             &      & \mcx{only cond. \ref{relevant_fp:unique} and \ref{relevant_fp:recurring}} & \mcx{characteristic} & \mcx{characteristic+manual}                                                                   \\
      No.    & Size & $|\mathcal{S}_c|$                                                         &                      & $|M_c|$                     & $|\mathcal{S}_c|$ &  & $|M_c|$ & $|\mathcal{S}_c|$ &  & $|M_c|$ \\
      \midrule
      1      & 38   & 3,752                                                                     &                      & 278,473                     & 3,282             &  & 131,842 & 3,232             &  & 70,228  \\
      2      & 36   & 1                                                                         &                      & 1                           & 1                 &  & 1       & 1                 &  & 1       \\
      3      & 14   & 40                                                                        &                      & 1,137                       & 34                &  & 694     & 2                 &  & 0       \\
      4      & 3    & 3                                                                         &                      & 3                           & 3                 &  & 3       & 3                 &  & 3       \\
      5      & 2    & 75                                                                        &                      & 4,191                       & 72                &  & 3,536   & 22                &  & 200     \\
      6      & 2    & 2                                                                         &                      & 81                          & 2                 &  & 81      & 1                 &  & 0       \\
      7      & 2    & 2                                                                         &                      & 0                           & 2                 &  & 0       & 2                 &  & 0       \\
      \midrule
      $\sum$ & 97   & 3,875                                                                     &                      & 283,887                     & 3,396             &  & 136,157 & 3,263             &  & 70,432  \\
      \bottomrule
    \end{tabularx}
  \end{table}


\subsection{Large Scale Evaluation}\label{sec:npm-scan}
  For large scale evaluation of our signatures, we harvested the npm repository on 25\textsuperscript{th} of September 2020.
  At this time 1,396,447 packages were listed and respectively 1,396,413 could be obtained in their \textquote{latest} version.
  In total 20,017,543 files and 749,558,178 function were inspected.

  On average, a npm package contains 15.6 files (min = 1, max = 82,530, $\sigma$ = 158.14) and an average package's size is 15.744 kB (min = 0 B, max = 145.7 MB, $\sigma$ = 200.82 kB).
  The average time needed for the transformation of a npm package into an \gls{ast} is 354.17 ms (min = 0 ms, max = 58 min, $\sigma$ = 5,611 ms).
  A corresponding \gls{ast} comprises 40.68 nodes on average (min = 1, max = 4,814,862, $\sigma$ = 1720.88).
  Overall, the experiment took around 48 hours and 154 GB of data was persisted in the database.

  \Cref{tab:cluster-sizes} also lists the number of matches ($|\mathcal{M}_c|$) our signatures produced per cluster.
  After the automated removal of false positive fingerprints, the total amount of matches went down from 283,887 to 136,157 (-52.04\%).
  For manual optimization we inspected the 50 most matching fingerprints for each cluster.
  This took roughly 10 minutes per cluster and resulted in 133 signatures (3.92\%) being removed.
  This further reduced the amount of matches from 136,157 to 70,432 (-48.27\%).

  It must be noted that the signatures of cluster 1 are responsible for most of the matches (99.71\%).
  This indicates that our approach fails to choose descriptive fingerprints from this cluster and hence produces many false positive matches.
  However, for the remaining clusters the \gls{acme} is able to pre-select a reasonable number of suspicious packages that need further manual inspection.

  In addition to automatically generated signatures, we manually created signatures for packages that did not belong to a cluster.
  To this end, we extracted the fingerprint of malicious functions by hand.
  This resulted in eight new pseudo-cluster with corresponding signatures.
  However, this yielded only one additional match.

\subsection{Detected Packages}
  By the construction of the signatures, see \Cref{sec:meth:signatures}, every match is treated as suspicious and hence needs manual inspection to verify actual maliciousness.
  Eventually, we were able to identify seven unreported but malicious packages that have code in common with known malicious packages.
  As listed in \Cref{tab:list-of-reported-malicious-packages}, we identified the packages \texttt{nodetest199}, \texttt{nodetest1010}, and \texttt{plutov-slack-client} based on the signature of cluster 4.

  The actual difference in source code between \texttt{nodetest199} and one of the clusters packages (\texttt{tensorplow}) from which the signature was derived is depicted in \Cref{fig:git-diff-nodetest199-tensorplow}.
  It is noticeable that \texttt{nodetest199} leverages almost the same code but with a modified address and port of the command and control server.
  However, all of these four packages open up a reverse shell when installed.
  We reported these packages to npm security and all of them received security advisories and were removed from the npm registry.

  \begin{figure}[tb]
    \begin{lstlisting}
--- a/nodetest199.js
+++ b/tensorplow.js
@@ -2,7 +2,7 @@
     var require = global.require || global.process.mainModule.constructor._load;
     if (!require) return;
     var cmd = (global.process.platform.match(/^win/i)) ? "cmd" : "/bin/sh";
<@\textcolor{red}{-    var net = require("net"),}@>
<@\textcolor{OliveGreen}{+    var net = require("tls"),}@>
         cp = require("child_process"),
         util = require("util"),
         sh = cp.spawn(cmd, []);
@@ -10,7 +10,9 @@
     var counter = 0;

     function StagerRepeat() {
<@\textcolor{red}{-        client.socket = net.connect(1111, "50.242.118.99", function() \{}@>
<@\textcolor{OliveGreen}{+        client.socket = net.connect(443, "45.63.54.27", \{}@>
<@\textcolor{OliveGreen}{+            rejectUnauthorized: false}@>
<@\textcolor{OliveGreen}{+        \}, function() \{}@>
             client.socket.pipe(sh.stdin);
             if (typeof util.pump === "undefined") {
                 sh.stdout.pipe(client.socket);
  \end{lstlisting}
    \caption{Difference of modified code fragments between tensorplow (known) and nodetest199 (detected by \gls{acme}). On line 8 a dependency was exchanges. Furthermore, the host address and port of the command and control server was changed on line 17. Nonetheless, \gls{acme} detected similarities based on mostly unchanged syntax around the modifications.}\label{fig:git-diff-nodetest199-tensorplow}
  \end{figure}

  Furthermore, \gls{acme} detected the packages \texttt{revshell} and \texttt{node-shells} that claim to be proof of concept packages.
  Thus, npm security did not publish a security advisory for \texttt{revshell} but nonetheless removed it from the registry.
  However, the package \texttt{node-shells} was published by Adam Baldwin, head of security at npm, and was thus not reported by us.

  The package \texttt{hellhun\_homelibrary} which was found by a manually generated signature was indeed not a malicious package itself.
  It is affected by \texttt{flatmap-stream}, the malicious package that was used in the \texttt{event-stream} incident.
  Hence, npm security decided to inform the developer about it instead of removing it.

  The publication dates of the packages related to cluster 4, namely \texttt{nodetest1010} (2018-08-03), \texttt{nodetest199} (2018-08-02), and \texttt{plutov-slack-client} (2018-03-30), fit the overall time frame of known malicious packages from that cluster (2018-03-06, 2018-09-08, 2018-03-25).
  Thus, we conclude that we identified remnants of a previous attack.

  However, the package \texttt{npmpubman} was published on 2020-09-13 which is way later than the average package from cluster 2.
  We reported it on 2020-09-28, only 15 days after it was published.

  \begin{table}[tb]
    \caption{List of reported malicious packages.}\label{tab:list-of-reported-malicious-packages}
    \begin{tabularx}{\textwidth}{Xrrr}
      \toprule
      Name                 & Cluster & Objective         & Advisory                                                  \\
      \midrule
      npmpubman            & 2       & data exfiltration & Yes, \href{https://www.npmjs.com/advisories/1568}{\#1568} \\
      plutov-slack-client  & 4       & reverse shell     & Yes, \href{https://www.npmjs.com/advisories/1569}{\#1569} \\
      nodetest1010         & 4       & reverse shell     & Yes, \href{https://www.npmjs.com/advisories/1570}{\#1570} \\
      nodetest199          & 4       & reverse shell     & Yes, \href{https://www.npmjs.com/advisories/1571}{\#1571} \\
      revshell             & 5       & reverse shell     & No, Proof of Concept (POC).                               \\
      node-shells          & 5       & reverse shell     & No, PoC                                                   \\
      hellhun\_homelibrary & manual  & financial gain    & No, affected by \texttt{flatmap-stream}.                  \\
      \bottomrule
    \end{tabularx}
  \end{table}

  Overall, we conclude that our automated generation of signatures reduces the workload for manual analysis drastically.
  However, manual optimization is still required to further boil down the amount of matches, thus further reducing the amount of suspicious packages for manual inspection.
  Nonetheless, our straight forward approach already yields feasible results.
  Hence, \gls{acme} can be leveraged to search packages for variations of known malicious code.
  In order to determine the suitability of application in real world, we discuss the scalability of our approach and possible limitations in the next section.

\section{Discussion}\label{sec:discussion}
  As shown in \Cref{sec:results}, our approach is able to reproduce the manual clustering of Ohm et al.\@~\cite{ohm2020backstabber}.
Thus, we are able to reduce the need for expert knowledge of detecting similar malicious packages by automatically selecting suspicious packages for further inspection.
Furthermore, the automatically generated and manually refined signatures can be used to efficiently scan a single package for variations of known malicious code fragments.
In particular, signatures were successfully used to perform a large scale analysis of all recent packages on npm.
Here, we identified seven unreported malicious packages leveraging off-the-shelf hardware.

\subsection{Efficiency}\label{sec:disc:efficiency}
  To demonstrate the theoretical efficiency of our approach, we discuss time and space complexity of all involved steps.
  For the initial generation of signatures, each known malicious package must be transferred into \gls{ast} representation.
  Assuming that most programming languages implement deterministic language to some extent this may be achieved in linear time\footnote{worst case $\mathcal{O}(n^3)$}~\cite{tomita1984lr,tomita1985efficient}.
  The creation of fingerprints depends on the employed hash functions which again takes linear time.
  Clustering of all malicious packages when using \gls{mcl} depends on the number of nodes $n$ in the graph as well as the parameter $k$ that keeps the matrix representation sparse.
  Hence, the time complexity of \gls{mcl} is $\mathcal{O}(n \cdot k^2)$~\cite{vandongen2000cluster}.
  Signatures are derived from identified clusters which again needs linear time.

  This procedure should be repeated periodically in order to keep clusters updated with known malicious packages that have been detected through other measures.
  As the dataset of malicious packages grows so does the detection capabilities of our approach.

  In order to match a new package against the signatures, it needs to be transformed into an \gls{ast} per source file and subsequently into a set of fingerprint.
  The actual time needed to match these fingerprints depends on the employed data storage.
  We leverage PostgreSQL which uses B-Trees and hence a lookup in $l$ elements takes $\mathcal{O}(\log{}l)$ in time.

  Considering space complexity, our approach needs to persist all \glsplural{ast}, corresponding fingerprints and results of the clustering.
  Each \gls{ast} contains at most all tokens in the source file and is hence linear in the amount of tokens.
  Each fingerprint is represented as a bytes-string of fixed length and thus constant in size.
  The database requires linear space to safe all fingerprints.
  Results of the clustering again depend on the number of nodes $n$ and the parameter $k$ which yields $\mathcal{O}(n \cdot k)$.

  Overall, both time and space complexity indicate good scalability and thus qualifying the approach for large scale deployment.

\subsection{Limitations}
  The use of \glspl{ast} and the leveraged abstraction level (c.f. \Cref{sec:detail}) are able to detect Type-1 and Type-2 clones by definition.
  By discarding identifier names, the approach becomes resilient against obfuscation through unreadable names and renaming in general.

  However, to fully support the detection of Type-3 clones, the comparison of two \glspl{ast} needs to be relaxed.
  One might leverage fuzzy hashes to allow similar but not exactly the same structure of code fragments.
  The detection of Type-4 clones -- semantic similarity -- is out of scope for an approach based on syntactically similarities.

  Furthermore, the detection of malicious packages based on signatures is solely able to detect variations of known malicious code fragments.
  Entirely new malicious source code will remain unnoticed by \gls{acme}.
  This, however, is a general downside of signature-based detection.
  Thus, \gls{acme} might be used in conjunction with anomaly-based detection.

  We evaluated \gls{acme} against a manual clustered dataset of known malicious packages.
  The manual clustering also took inter dependency into account while we only focused on syntactic similarities.
  However, only few clusters based on dependency exists and hence \gls{acme} was able to reproduce the manual results almost perfectly.
  If the manual clustering is flawed, our approach also contains these flaws.

  \gls{acme} performed poorly for packages of cluster 1 (c.f. \Cref{tab:cluster-sizes}) which is responsible for most of the matches.
  This indicates that a more sophisticated selection of relevant fingerprints might be needed.
  Solely 204 suspicious packages need manual inspection when leaving out matches from that cluster.

  However, fingerprints causing false positives may need to be sorted out by hand.
  By removing the 50 most matching false positive fingerprints from each cluster we reduced the amount of matches by roughly 50\%.
  The manual process of eradicating false positive fingerprints is cumbersome but with about 10 minutes per 50 fingerprints still feasible.
  Nonetheless, there is still need for manual inspection that might be canceled out through a more sophisticated signature generation.

  \glsresetall

\section{Conclusion}\label{sec:conclusion}
  In this paper, we examined how source code similarities of known malicious packages can be leveraged to support the detection of software supply chain attacks.
Our main goal is to aid analysts that commonly detect malicious software packages that are distributed by large package repositories like npm.
To this end, we automatized the clustering and signature generation of known malicious packages which is typically based on expert knowledge and manual inspection.
On a dataset of 114 malicious npm packages that have been used in real world attacks, we evaluated several approaches to find and group syntactical similarities in source codes.
Based on that, clusters of packages with similar structure were identified automatically and unsupervised.

Compared to the manual clustering of these packages at hand, our best approach yields promising results ($F_{1}=0.99$).
It leverages \glspl{ast} to compare source code of multiple packages and \gls{mcl} to identify clusters among these and is hence called \gls{acme}.

In conclusion, we are able to systematize and automatize the unsupervised detection of related malicious packages.
Our approach can be used to support analysts by pre-selecting suspicious packages based on signatures of known malicious code fragments for manual inspection.
This reduces the need for expert knowledge and manual inspection drastically.

Eventually, \gls{acme} identified seven clusters in the leveraged dataset.
Subsequently, signatures that characterize malicious packages from a particular cluster were derived.
In order to minimize false positives, we removed parts of the signatures that matched on the 108 most depended upon packages from npm.

To demonstrate the effectiveness of \gls{acme}, a scan of the whole npm registry based on all generated signatures was performed.
This revealed seven previously unreported packages in total.
A manual inspection showed that four of them are indeed malicious packages and were therefore reported by us and subsequently removed from npm.
Two of the remaining three were proof of concept packages.
The last package itself is not malicious but it contained a full copy of the known malicious package \texttt{flatmap-stream} as dependency.

In conclusion this means that our approach is feasible to automatically generate signatures for known malicious packages which then may be used to scan packages for known malicious code.
Packages detected this way can be presented to an analyst which may verify the package's maliciousness.
Through the use of \glspl{ast} the approach is resilient to modifications of the source code like renaming of variables (Type-1 and Type-2 clones) and minor structural modifications (Type-3 clones).
Furthermore, \gls{acme} is transferable to any other programming language.
However, automatically generated signatures may not yet be perfect as they still may cause false positives which may be removed manually.
Nonetheless, our naive approach already yields promising results and good scalability.

For future work we plan to optimize our signature generation and enhance support for structural modifications (Type-3 clones).
Eventually, we would like to expand our approach to other software ecosystems like \gls{pypi} and RubyGems.

  \subsubsection*{Acknowlegements.}
    This work is funded under the SPARTA project, which has received funding from the European Union's Horizon 2020 research and innovation programme under grant agreement No 830892.

    \bibliographystyle{splncs04}
    \bibliography{bibliography.bib}

\begin{thebibliography}{10}
\providecommand{\url}[1]{\texttt{#1}}
\providecommand{\urlprefix}{URL }
\providecommand{\doi}[1]{https://doi.org/#1}

\bibitem{allard20markov}
Allard, G.: Markov clustering (2020),
  \url{https://github.com/GuyAllard/markov_clustering}

\bibitem{bilgin2020vulnerability}
Bilgin, Z., Ersoy, M.A., Soykan, E.U., Tomur, E., {\c{C}}omak, P.,
  Kara{\c{c}}ay, L.: Vulnerability prediction from source code using machine
  learning. IEEE Access  \textbf{8},  150672--150684 (2020)

\bibitem{carnogursky2019attacks}
{\v{C}}arnogursk{\`y}, M.: Attacks on Package Managers. Master's thesis,
  Masaryk University, Faculty of Informatics (2019)

\bibitem{chilowicz2009syntax}
Chilowicz, M., Duris, E., Roussel, G.: Syntax tree fingerprinting for source
  code similarity detection. In: 2009 IEEE 17th International Conference on
  Program Comprehension. pp. 243--247. IEEE (2009)

\bibitem{chinthanet2020code}
Chinthanet, B., Ponta, S.E., Plate, H., Sabetta, A., Kula, R.G., Ishio, T.,
  Matsumoto, K.: Code-based vulnerability detection in node. js applications:
  How far are we? arXiv preprint arXiv:2008.04568  (2020)

\bibitem{cosma2011approach}
Cosma, G., Joy, M.: An approach to source-code plagiarism detection and
  investigation using latent semantic analysis. IEEE transactions on computers
  \textbf{61}(3),  379--394 (2011)

\bibitem{djuric2013source}
Djuri{\'c}, Z., Ga{\v{s}}evi{\'c}, D.: A source code similarity system for
  plagiarism detection. The Computer Journal  \textbf{56}(1),  70--86 (2013)

\bibitem{duan2020measuring}
Duan, R., Alrawi, O., Kasturi, R.P., Elder, R., Saltaformaggio, B., Lee, W.:
  Measuring and preventing supply chain attacks on package managers. arXiv
  preprint arXiv:2002.01139  (2020)

\bibitem{garrett2019detecting}
Garrett, K., Ferreira, G., Jia, L., Sunshine, J., K{ä}stner, C.: Detecting
  suspicious package updates. In: 2019 IEEE/ACM 41st International Conference
  on Software Engineering: New Ideas and Emerging Results (ICSE-NIER). pp.
  13--16. IEEE

\bibitem{hagberg2008exploring}
Hagberg, A., Swart, P., S~Chult, D.: Exploring network structure, dynamics, and
  function using networkx. Tech. rep., Los Alamos National Lab.(LANL), Los
  Alamos, NM (United States) (2008)

\bibitem{henderson2013zhang}
Henderson, T., Johnson, S.: Zhang-shasha: Tree edit distance in python,
  \url{https://pythonhosted.org/zss}

\bibitem{kruskal1983overview}
Kruskal, J.B.: An overview of sequence comparison: Time warps, string edits,
  and macromolecules. SIAM review  \textbf{25}(2),  201--237 (1983)

\bibitem{li2016vulpecker}
Li, Z., Zou, D., Xu, S., Jin, H., Qi, H., Hu, J.: Vulpecker: an automated
  vulnerability detection system based on code similarity analysis. In:
  Proceedings of the 32nd Annual Conference on Computer Security Applications.
  pp. 201--213 (2016)

\bibitem{liu2006gplag}
Liu, C., Chen, C., Han, J., Yu, P.S.: Gplag: detection of software plagiarism
  by program dependence graph analysis. In: Proceedings of the 12th ACM SIGKDD
  international conference on Knowledge discovery and data mining. pp. 872--881
  (2006)

\bibitem{haverbeke2020acorn}
Marijn~Haverbeke, I.S., et~al.: Acorn (2020),
  \url{https://github.com/acornjs/acorn}

\bibitem{mcinnes2017hdbscan}
McInnes, L., Healy, J., Astels, S.: hdbscan: Hierarchical density based
  clustering. Journal of Open Source Software  \textbf{2}(11), ~205 (2017)

\bibitem{novak2019source}
Novak, M., Joy, M., Kermek, D.: Source-code similarity detection and detection
  tools used in academia: a systematic review. ACM Transactions on Computing
  Education (TOCE)  \textbf{19}(3),  1--37 (2019)

\bibitem{ohm2020backstabber}
Ohm, M., Plate, H., Sykosch, A., Meier, M.: Backstabber's knife collection: A
  review of open source software supply chain attacks. In: International
  Conference on Detection of Intrusions and Malware, and Vulnerability
  Assessment. Springer (2020)

\bibitem{ohm2020towards}
Ohm, M., Sykosch, A., Meier, M.: Towards detection of software supply chain
  attacks by forensic artifacts. In: Proceedings of the 15th International
  Conference on Availability, Reliability and Security. pp.~1--6. ACM (2020)

\bibitem{pedregosa2011scikit}
Pedregosa, F., Varoquaux, G., Gramfort, A., Michel, V., Thirion, B., Grisel,
  O., Blondel, M., Prettenhofer, P., Weiss, R., Dubourg, V., et~al.:
  Scikit-learn: Machine learning in python. the Journal of machine Learning
  research  \textbf{12},  2825--2830 (2011)

\bibitem{pfretzschner2017identification}
Pfretzschner, B., ben Othmane, L.: Identification of dependency-based attacks
  on node. js. In: Proceedings of the 12th International Conference on
  Availability, Reliability and Security. pp.~1--6 (2017)

\bibitem{ragkhitwetsagul2018comparison}
Ragkhitwetsagul, C., Krinke, J., Clark, D.: A comparison of code similarity
  analysers. Empirical Software Engineering  \textbf{23}(4),  2464--2519 (2018)

\bibitem{seargeek11fuzzywuzzy}
seatgeek: Fuzzywuzzy: Fuzzy string matching in python (2011),
  \url{https://chairnerd.seatgeek.com/fuzzywuzzy-fuzzy-string-matching-in-python/}

\bibitem{taylor2020spellbound}
Taylor, M., Vaidya, R.K., Davidson, D., De~Carli, L., Rastogi, V.: Spellbound:
  Defending against package typosquatting

\bibitem{tomita1984lr}
Tomita, M.: Lr parsers for natural languages. In: 10th International Conference
  on Computational Linguistics and 22nd Annual Meeting of the Association for
  Computational Linguistics. pp. 354--357 (1984)

\bibitem{tomita1985efficient}
Tomita, M.: An efficient context-free parsing algorithm for natural languages.
  In: IJCAI. vol.~2, pp. 756--764. Citeseer (1985)

\bibitem{tschacher2016typosquatting}
Tschacher, N.P.: Typosquatting in programming language package managers

\bibitem{vandongen2000cluster}
vanDongen, S.: A cluster algorithm for graphs. Information Systems [INS] (R
  0010) (2000)

\bibitem{vu2020typosquatting}
Vu, D.L., Pashchenko, I., Massacci, F., Plate, H., Sabetta, A.: Typosquatting
  and combosquatting attacks on the python ecosystem

\bibitem{walker2020open}
Walker, A., Cerny, T., Song, E.: Open-source tools and benchmarks for
  code-clone detection: past, present, and future trends. ACM SIGAPP Applied
  Computing Review  \textbf{19}(4),  28--39 (2020)

\bibitem{yamaguchi2012generalized}
Yamaguchi, F., Lottmann, M., Rieck, K.: Generalized vulnerability extrapolation
  using abstract syntax trees. In: Proceedings of the 28th Annual Computer
  Security Applications Conference. pp. 359--368 (2012)

\bibitem{zhang1989simple}
Zhang, K., Shasha, D.: Simple fast algorithms for the editing distance between
  trees and related problems. SIAM journal on computing  \textbf{18}(6),
  1245--1262 (1989)

\end{thebibliography}

\end{document}